\documentclass[referee]{mn2e}
\usepackage{psfig}

\def\ltsima{$\; \buildrel < \over \sim \;$}
\def\lsim{\lower.5ex\hbox{\ltsima}}
\def\gtsima{$\; \buildrel > \over \sim \;$}
\def\gsim{\lower.5ex\hbox{\gtsima}}

\begin{document}
\title[Global solution of accretion disk around rotating compact objects]
{Global solution of viscous accretion disk around rotating compact objects: a pseudo-general-relativistic study}

\author[ Mukhopadhyay \& Ghosh]
{Banibrata Mukhopadhyay$^{1}$ \& Shubhrangshu Ghosh$^{2}$\\
$^1$ Inter-University Centre for Astronomy and Astrophysics, Post Bag 4, Ganeshkhind,
Pune-411007, India\\ $^2$ Department of Physics and Center for Theoretical Studies, Indian Institute of
Technology, Kharagpur-721302, India\\ 
 }

\maketitle

\noindent{Accepred for publication in {\it Monthly Notices of the Royal Astronomical Society}}

\begin{abstract}

We study the solution of viscous accretion disks around rotating compact/central
object having hard surface i.e. neutron star, strange star and any other highly 
gravitating objects. We choose pseudo-Newtonian approach to describe the
relativistic accretion disk. For this purpose, a new pseudo-Newtonian potential is established
which is applicable to describe the relativistic properties of star and its disk.
As we know, the Hartle-Thorne metric can describe geometry of star as well as the space-time
out-side of it, we use this metric to establish our potential. Our potential reproduces
the marginally stable orbit exactly as that of general relativity. It also reproduces the
marginally bound orbit and specific mechanical energy at the marginally stable orbit 
with at most $4\%$ and $10\%$ error respectively. Using this potential we study the global
parameter space of the accretion disk. Thus, we find out the physical parameter regime, 
for which the stable accretion disk can be formed around gravitating object with hard
surface. We also study, how the fluid properties get changed with different rotations of the 
central star. We show that with the change of rotation to the central object, the valid disk 
parameter region dramatically changes. We also show the effect of viscosity to the 
fluid properties of the disk. Subsequently, we give a theoretical prediction of kHz QPO,
at least for one out of a pair, for a fast rotating compact object as 4U 1636-53.

\end{abstract}


\section{Introduction}


For over last three decades, fluid dynamical study of accretion disks around compact
object is indeed a topic extensively discussed, mostly for black holes 
(Abramowicz et al. 1988, 1996; Abramowicz \& Kato 1989;
Chakrabarti 1990, 1996a,b,c; Narayan \& Yi 1994; Narayan et al.
1997, 1998; Kato et al. 1998; Sponholz \& Molteni 1994;
Peitz \& Appl 1997; Gammie \& Popham 1998; Popham \& Gammie 1998; Miwa et al. 1998; 
Lu \& Yuan 1998; Manmoto 2000) and a few for neutron stars
(Chakrabarti \& Sahu 1997; Mukhopadhyay 2000, 2002a; Popham \& Sunyaev 2001). 
Though in some of the cases authors use full general relativistic approach, in most of the cases
the approaches are pseudo-Newtonian. In the pseudo-Newtonian method, they use simple
non-relativistic equations to describe the disk with the appropriate choice of
modified gravitational force which takes care about the geometry of back-ground space-time.
There are a number of basic differences between the accretion disk around black hole
and any other gravitating objects namely neutron star, white dwarf, strange star and 
other gravitating stars particularly at the inner edge of the disk. 
(1) While the black hole has event horizon, other gravitating objects mentioned above have
hard surface. Thus, whereas in the case of other gravitating object close to the surface 
the matter speed must be subsonic, for black holes the matter speed is supersonic.
(2) In case of other gravitating object that has hard surface, the matter comes into contact with
its outer surface. Thus, there is an obvious effect of dragging to the exterior space-time
due to its rotation. In terms of general relativity, corresponding space-time metric 
should be the solution for interior as well as exterior to it. Therefore, it is 
necessary to consider the 'dragging effect' of rotating gravitating star (particularly for
neutron star, where the relativistic effect is very important) to the space-time.
In case of black hole, which does not have hard surface, this effect comes to the space-time 
automatically through the 'Lense-Thirring drag' from the Kerr geometry which does not 
describe any physics interior to the horizon. 
(3) Further, the overall temperature as well as the luminosity of the
disk is expected to be high in the cases of other gravitating star particularly for
neutron star with respect to that of black hole (Mukhopadhyay 2000, 2002a) for a particular
accretion rate. If the same parameters are chosen for the infalling matter towards 
black hole and other gravitating objects separately, at same accretion rate, the maximum 
temperature in the disk around black hole will be lower at least a few factors times. 
Thus, in the disk around neutron star, expected generation/absorption of nuclear energy
is higher than that of black hole (Mukhopadhyay 2002a). As there is a clear distinction
between accretion disks around black hole and other gravitating objects, hereinafter, we will
call the class of gravitating objects having hard surface as GOHS to distinguish from
black hole. GOHS can be meant as neutron star, strange star, white dwarf and any other
highly gravitating objects where disk may form. As the radius of white dwarf
is very large, the general relativity is not much important in the accretion disk around it.
But our discussions will be generally true for disk around any kind of GOHS.

Mukhopadhyay (2002a) has studied the accretion disk around 
weakly magnetized slowly rotating neutron stars, and found that the
temperature of the inner disk is significantly higher than the case of
black holes. Prasanna \& Mukhopadhyay (2003) have described the accretion disk
with the inclusion of 'Coriolis' effect. They have shown that the rotational effect
of compact object to the accretion disk can be approximately reproduced with the use 
of Coriolis effect. They have done perturbative analysis and shown that the 
{\it self-similar solutions} are well behaved for most of the parameter space. 
As no cosmic object is static, in this paper, 
our interest is to include the rotation of the neutron star, globally speaking rotation of 
GOHS, in a more proper manner and see its effect onto the disk. 
The recent observation tells about the evidence of millisecond pulser 
(Strohmayer \& Markwardt 2002) of frequency 582Hz for 4U 1636-53 which is the accreting LMXB. 
Therefore, it is important to consider the rotational effect of the GOHS to the disk.
We approach our study in pseudo-Newtonian manner. So far, there is no pseudo-potential in the 
literature which can describe the accretion disk around GOHS. Thus, first of all we
should establish such a potential useful for our work.

Shakura \& Sunyaev (1973) first initiated the accretion disk modeling using simple Newtonian 
potential meant for non-rotating black holes. As the relativistic effects are extremely important 
near the compact objects, this potential can not describe the essential inner properties of the disk.
Paczy\'nski \& Wiita (1980) modified this Newtonian potential in conformity with the  
Schwarzschild geometry, which can naturally reveal approximately the properties of disk around
non-rotating black holes and GOHSs, even that of the inner most part of the disk. 
The potential can reproduce the radius of marginally stable orbit $(x_s)$ and
marginally bound orbit $(x_b)$, mechanical energy per unit mass at the last 
stable circular orbit $(E_s)$ and viscous efficiency of the disk.
The last two parameters agree within a $10\%$ error but the first two
cases exactly match with that of the Schwarzschild metric. 
Nowak \& Wagoner (1991) proposed another potential for accretion disk 
around non-rotating black holes. This potential can also mimic approximately most of the properties of 
the disk governed by Schwarzschild geometry. However, angular and epicyclic frequency of the 
disk can be best analyzed with this potential.
Artemova et al. (1996) proposed a couple of pseudo-potentials to describe accretion disk
for rotating and non-rotating black holes. These potentials are well analyzed
by a number of astrophysicists later (e.g. Lovas 1998; Semer\'ak \& Karas 1999). Very recently, 
Mukhopadhyay (2002b) and Mukhopadhyay \& Misra (2003) prescribed some new potentials with which properties of accretion disk can be very well described in Kerr geometry. The more interesting 
fact lies in the methodology adopted by Mukhopadhyay (2002b) to describe a potential in 
the accretion disk, which can be used to derive the pseudo-potential for any metric 
according to the physics concerned.
The potentials proposed by Mukhopadhyay \& Misra (2003) can be used for 
the time-dependent simulation of accretion disk. But all the potentials are described either for
the fluid dynamical study of disk around black hole or the study of temporal effect around
compact objects. As the fluid properties of inner accretion disk are very much influenced 
by the nature of central gravitating object as mentioned above, it is essential to establish
a pseudo-potential which can describe an accretion disk around GOHS. Although in case of black 
hole, the space-time can be described by Kerr geometry, but in the case of GOHS with slow rotation,
the corresponding metric which can continuously describe the space-time inside and outside the
star is Hartle-Thorne (hereinafter HT, Hartle \& Thorne 1968). Thus, we should establish our
potential according to HT metric.

Thus, we shall consider the general solutions for fluid disks around rotating GOHS.
In order to consider a solution procedure, we will follow Chakrabarti (1996b) where
the global solution for transonic accretion flows is studied around rotating black holes
but in the case of weak viscosity. Here, we will consider the set of disk equations in terms of
general viscous prescription and the accretion around GOHS. 
As we want to concentrate upon the features of the inner region, we
shall restrict the study to 'sub-Keplerian' flow. We study, how does the rotation of GOHS
affect on the global parameter space of accretion disk. We know that for the case of
non-rotating compact object, the stable accretion disk forms only for a certain set of
physical parameter. We would like to check how this parameter region gets affected,
how does the steady disk structure change, with different values of the rotation to GOHS.
Once we have the clear picture about the parameter space, we will concentrate upon the
fluid properties of the disk for different viscosity and how does the viscous fluid properties
change with different rotations of GOHS. 

Therefore, we are going to present a complete set of work on accretion disk around GOHS.
In order to do so, in the next section, we will derive the required pseudo-Newtonian potential
and establish its efficiency.
In \S 3, we will present the basic equations of viscous accretion disk around GOHS. In \S 4 and \S 5,
we will analyse the parameter space of the accretion disk around GOHS and corresponding
properties of accreting fluid respectively.
Finally we shall present our discussion and overall summary in \S 6.

\section{Pseudo-Newtonian potential and its efficiency}

Here, following Mukhopadhyay (2002b), we will derive the pseudo-Newtonian potential for 
accretion disk modeling around GOHS according to HT metric.  
The Lagrangian density for a particle in the HT geometry (neglecting
the higher quadruple terms) at the equatorial plane ($\theta=\pi/2$) can be written as
\begin{eqnarray}
2{\cal L}=-\left(1-\frac{2M}{r}-\frac{2J^2}{r^4}\right){\dot 
t}^2+\left(1-\frac{2M}{r}+\frac{2J^2}{r^4}\right)^{-1}{\dot r}^2
+r^2{\dot \phi}^2-\frac{4J}{r}{\dot \phi}{\dot t},
\label{lag}
\end{eqnarray}
where over-dots denote the derivative with respect to the proper-time 
$\tau$ and $J$ denotes the angular momentum of GOHS. As 
the metric is valid only for slowly rotating stars, the rotation parameter $J$ is 
restricted as $J\leq 0.5$.
The geodesic equations of motion are 
\begin{eqnarray}
E={\rm constant}=\left(1-\frac{2M}{r}-\frac{2J^2}{r^4}\right){\dot 
t}+\frac{2J}{r}{\dot \phi},
\label{en}
\end{eqnarray}
\begin{eqnarray}
\lambda={\rm constant}=r^2{\dot \phi}-\frac{2J}{r}{\dot t}.
\label{ang}
\end{eqnarray}
For the particle with non-zero rest mass, $g_{\mu\nu}p^\mu p^\nu=-m^2$ 
(where $p^\mu$ is the momentum of the particles and $g_{\mu\nu}$ is the metric). 
Replacing the solution for $\dot t$ and $\dot \phi$ from ({\ref{en}) and (\ref{ang}) into 
(\ref{lag}) gives a differential equation for $r$ 
\begin{eqnarray}
\nonumber
\left(\frac{dr}{d\tau}\right)^2&=&\left(\frac{4h(r)J^2}{g(r)r^6}
+\frac{16J^4}{g(r)r^{10}}-\frac{g(r)}{r^2}\right)\lambda^2+\left(\frac{h(r)}{g(r)}
+\frac{4J^2}{g(r)r^4}\right)E^2 \\ 
&+&\left(-\frac{4h(r)J}{g(r)r^3}-\frac{16J^3}{g(r)r^7}\right)E\lambda-g(r)m^2=\Psi,
\label{dr}
\end{eqnarray}
where, $g(r)=\left(1-\frac{2m}{r}+\frac{2J^2}{r^4}\right)$, $h(r)=
\left(1-\frac{2m}{r}-\frac{2J^2}{r^4}\right)$.
Here, $\Psi$ can be identified as an effective potential for the radial 
geodesic motion.
The conditions for circular orbits are
\begin{eqnarray}
\Psi=0,\hskip1.cm\frac{d\Psi}{dr}=0.
\label{psi}
\end{eqnarray}
Solving for $E$ and $\lambda$ from (\ref{psi}) we get
\begin{eqnarray}
E=X(r),
\label{enf}
\end{eqnarray}
\begin{eqnarray}
\nonumber
\lambda &=&X(r)\sqrt{\frac{m^4 r^6(-4J^2+Mr^3)^2}{[-2J^3 m^2+Jm^2 r^3(-4m+3r)+\sqrt{m^4(5J^2+Mr^3)
(2J^2+r^3(-2M+r))^2}]^2}}, \\
\label{angf}
\end{eqnarray}
where,
\begin{eqnarray}
\nonumber
X(r)&=&\frac{1}{m^2r^3(-4J^2+Mr^3)}\left[2J^3m^2+Jm^2(4M-3r)r^3\right. \\
\nonumber
&&\left . -\sqrt{m^4(5J^2+Mr^3)(2J^2+r^3(-2M+r))^2}\right]\\
\nonumber
&&\{(-12J^4m^2r^2+2J^2m^2(9M-7r)r^5+m^2M(3M-r)r^8 \\
\nonumber
&+&6Jr^2\sqrt{m^4(5J^2+Mr^3)(2J^2+r^3(-2M+r))^2})/ \\
&&(36J^4-r^6(-3M+r)^2+12J^2r^3(-3M+2r))\}^{1/2}. 
\label{enf}
\end{eqnarray}
Now as standard practice, we can define the Keplerian angular momentum distribution, 
$\lambda_K=\frac{\lambda}{E}$. Therefore, corresponding centrifugal force in HT geometry can be 
written as
\begin{eqnarray}
\frac{\lambda_K^2}{r^3}=\frac{m^4 r^3(-4J^2+Mr^3)^2}{[-2J^3 m^2
+Jm^2 r^3(-4m+3r)+\sqrt{m^4(5J^2+Mr^3)(2J^2+r^3(-2M+r))^2}]^2}=F_r.
\label{lamkep}
\end{eqnarray}
Thus from above, $F_r$ can be identified as the gravitational force of 
rotating GOHS at the Keplerian orbit in an equatorial plane.
If we choose $x=r/M$ (as $G=c=1$) and $m=1$, above expression reduces to
\begin{eqnarray}
F=\frac{ x^3(-4J^2+x^3)^2}{[-2J^3 +J x^3(-4+3x)+\sqrt{(5J^2+x^3)(2J^2+x^3(-2+x))^2}]^2}.
\label{lamkep1}
\end{eqnarray}
Also the above expression reduces to Paczy\'nski-Wiita (1980) form for $J=0$.
Thus we propose that (\ref{lamkep1}) is the most general form of the
gravitational force corresponding to the pseudo-potential in accretion 
disk around GOHS at equatorial plane.

\subsection{Comparison of the Results for Hartle-Thorne geometry and 
Pseudo-Potential}

Now we will establish the validity of this potential. We would like to check, whether
this potential can reproduce the values of marginally bound ($x_b$), marginally stable ($x_s$)
orbits and mechanical energy per unit mass at $x_s$ ($E_s$), as same as HT geometry or not. 

In terms of the peudo-potential, one can write the radial velocity ($v$), angular
velocity ($\Omega$) and angular momentum per unit mass ($\lambda$) for
a Keplerian orbit in an accretion disk as 
(Lynden-Bell 1969, Pringle \& Rees 1973, Shakura \& Sunyaev 1973, Novikov \& Thorne 1973, 
Paczy\'nski \& Wiita 1980),
\begin{eqnarray}
v=\sqrt{x\frac{dV}{dx}},\hskip1cm\Omega=\sqrt{\frac{1}{x}\frac{dV}{dx}},\hskip1cm
\lambda=\sqrt{x^3\frac{dV}{dx}}.
\label{vol}
\end{eqnarray}
At the marginally bound orbit, mechanical energy $E$ reduces to zero and we get
\begin{eqnarray}
\frac{v^2}{2}+V=\frac{x}{2}\frac{dV}{dx}+V=0,
\label{rb}
\end{eqnarray}
where, $V=\int F dx$, which can be used to calculate $x_b$ for the potential.
For the last stable orbit ($x_s$), $d\lambda/dx=0$, for our pseudo-potential which gives
\begin{eqnarray}
\nonumber
x^5\left[-8 J^4+2J^2(5-2x)x^3+(x-2)x^6\right]\left[-240J^6+(x-6)x^9+8J^4x^3(12+5x)\right.\\
\left.+2J^2x^6(20x-33)+(24J^3+12Jx^3)\sqrt{(5J^2+x^3)(2J^2+(x-2)x^3)^2}\right]=0.
\label{rs}
\end{eqnarray}
The solution of (\ref{rs}) gives the location of the last stable circular orbit ($x_s$).
For any $J$, the value of $x_s$ computed from the above equation (\ref{rs}),
matches exactly with the radius of last stable circular
orbit in HT geometry. In Table-1, 2 and 3, we list $x_b$, $x_s$
and $E_s$ respectively for the potential $V$ and HT geometry for various values of $J$.

\clearpage
\vskip0.2cm
{\centerline{\large Table-1}}
{\centerline{\large Values of $x_b$}}
\begin{center}
{
\vbox{
\begin{tabular}{llllllllllllllllllllllllllllllll}
\hline
\hline
 $J$ & $0$ &$0.1$ & $0.2$ & $0.3$ & $0.4$ & $0.5$    \\
\hline
\hline
 $V$  & $4$ & $3.7862$ & $3.5619$ & $3.3255$ & $3.0762$ &  $2.8135$ 
\\
\hline
 HT & $4$ & $3.7954$ & $3.5810$ & $3.3557$ & $3.118$ & $2.8689$   \\
\hline
\hline
 $J$ & $0$  & $-0.1$& $-0.2$ & $-0.3$ & $-0.4$ & $-0.5$   \\
\hline
\hline
 $V$  & $4$&  $4.2042$ & $4.4$ & $4.5885$ & $4.77$ & $4.9462$  
\\
\hline
 HT & $4$ & $4.1957$& $4.3838$ & $4.5684$ & $4.7397$ & $4.9089$   \\
\hline
\hline
\end{tabular}
}}
\end{center}

\vskip0.2cm
{\centerline{\large Table-2}}
{\centerline{\large Values of $x_s$}}
\begin{center}
{
\vbox{
\begin{tabular}{llllllllllllllllllllllllllllllll}
\hline
\hline
 $J$ & $0$ &$0.1$ &$0.2$&  $0.3$ &$0.4$ & $0.5$    \\
\hline
\hline
$V$ \& HT  & $6.0$ & $5.6648$& $5.3107$ & $4.9339$ & $4.5299$ & $4.0934$    \\
\hline
\hline
 $J$ & $0$  & $-0.1$& $-0.2$ & $-0.3$ & $-0.4$ & $-0.5$  \\
\hline
\hline
$V$ \& HT & $6.0$ & $6.3189$ & $6.6240$ & $6.9170$  & $7.1994$ & $7.4724$   \\
\hline
\hline
\end{tabular}
}}
\end{center}
\vskip0.5cm

{\centerline{\large Table-3}}
{\centerline{\large Values of $E_s$}}
\begin{center}
{
\vbox{
\begin{tabular}{lllllllllllllllllll}
\hline
\hline
 $J$ & $0$ &$0.1$ & $0.2$ & $0.3$ & $0.4$ & $0.5$    \\
\hline
\hline
 $V$  & $-0.0625$ &$-0.0664$ & $-0.0711$ & $-0.0769$ & $-0.0843$ & $-0.0943$ \\
\hline
 HT & $-0.0572$ & $-0.0607$& $-0.0649$ &  $-0.0701$ & $-0.0767$ & $-0.0857$ \\
\hline
\hline
 $J$ & $0$  & $-0.1$ & $-0.2$ & $-0.3$ &$-0.4$ &$-0.5$   \\
\hline
\hline
 $V$ & $-0.0625$ & $-0.0592$  & $-0.0563$ & $-0.0538$ & $-0.0516$ & $-0.0497$ \\
\hline
 HT & $-0.0572$ & $-0.0542$ & $-0.0517$ & $-0.0494$ & $-0.0474$ & $-0.0457$\\ 
\hline
\hline
\end{tabular}
}}
\end{center}

From Table-1, it is clear that for all values of $J$, $V$ can 
reproduce the value of $x_b$ in very good
agreement with general relativistic (HT) results. The maximum error in 
$x_b$ 
is $\sim 4\%$.  Table-2 indicates that $V$ reproduces $x_s$ exactly
as that of HT geometry. Table-3 shows a maximum possible error in 
$E_s$ is $\sim 10\%$. Thus, the potential $V$ will produce a slightly
larger luminosity than the general relativistic one in the accretion 
disk for a particular value of $J$. It is to be noticed that for 
counter-rotating GOHS, that is for retrograde orbits, the errors are less 
than those of co-rotating ones. Eventually, we can tell that our 
potential can describe approximately all the phenomena
of that of HT geometry and can claim it as a good  pseudo-potential 
to describe the relativistic accretion disk around the rotating GOHS,
particularly close to the equatorial plane. Thus, for a thin disk or
vertically averaged disk, this is an ideal pseudo-potential.

\section{Basic equations of accretion disk and their consequences }\label{sec:ref}

Here, throughout our calculations, we express the radial coordinate
in unit of $GM/c^2$, where $M$ is mass of the compact object, $G$ is the gravitational constant and
$c$ is the speed of light. We also express the velocity in unit of speed of
light and the angular momentum in unit of $GM/c$. The equations to be solved are given below.\\
\noindent (1) Equation of continuity:
   \begin{equation}
   -4\pi x\Sigma v=\dot{M},
   \label{ec}
   \end{equation}
   where $\Sigma$ is the vertically integrated density and $\dot{M}$ is the accretion rate. 
Following Matsumoto et al. (1984), we can calculate the $\Sigma$ as
\begin{equation}
  \Sigma=I_n\rho_e h(x), 
  \label{den}
  \end{equation}
  where $\rho_e$ is the density at equatorial plane, $h(x)$ is the half-thickness of the disk and
  $I_n=\frac{(2^n n!)^2}{(2n+1)!}$, where $n$ is the polytropic index. From the vertical equilibrium 
  assumption, the half-thickness can be written as
  \begin{equation}
   h(x)=c_s x^{1/2}F^{-1/2},
  \label{ht}
  \end{equation}
  where $c_s$ is the speed of sound.\\ 
  \noindent (2) Radial momentum balance equation:
   \begin{equation}
   v\frac{dv}{dx}+\frac{1}{\rho}\frac{dP}{dx}-\frac{\lambda^2}{x^3}+F(x)=0, 
   \label{rmom}
   \end{equation}
   where the flow is chosen adiabatic with the equation of state to be $P=k\rho^\gamma$ and $c_s^2=\frac{\gamma P}{\rho}$.\\
   \noindent (3) Azimuthal momentum balance equation:
   Here, we will follow Chakrabarti (1996a) to express the viscous dissipation $Q^+$ in terms of shear stress
   $W_{x\phi}$. Because of this viscous dissipation, the angular momentum varies in accretion disk.
   Thus, $W_{x\phi}=-\alpha(I_{n+1}P+I_n v^2\rho)h(x)$ and $Q^+=\frac{W_{x\phi}^2}{\eta}$, $\eta$ is 
   coefficient of viscosity and $\alpha$ is Shakura-Sunyaev (1973) viscosity parameter. Thus the equation to be
   \begin{equation}
   v\frac{d\lambda}{dx}=\frac{1}{\rho h(x) x}\frac{d}{dx}\left[x^2\alpha\left(\frac{I_{n+1}}{I_n}P+v^2\rho\right)h(x)\right].
   \label{azmom}
   \end{equation}
(4) Entropy equation: According to the mixed shear stress (Chakrabarti 1996a), 
$Q^+=-\alpha(I_{n+1}P+I_n v^2\rho)h(x)x\frac{d\Omega}{dx}$. For simplicity, we also consider the heat lost
is proportional to the heat gained by the flow. Thus the equation to be
   \begin{equation}
   \Sigma vT\frac{ds}{dx}=\frac{vh(x)}{\Gamma_3-1}\left(\frac{dP}{dx}-\Gamma_1\frac{P}{\rho}\frac{d\rho}{dx}\right)=Q^+-Q^-=fQ^+,
   \label{enteq}
   \end{equation}
  where $s$ is entropy density and $f$ is cooling factor which is close to $0$ and $1$ for the flow with efficient 
  and inefficient cooling respectively. Following Cox \& Giuli (1968), we can define
  \begin{equation}
  \nonumber
   \Gamma_3=1+\frac{\Gamma_1-\beta}{4-3\beta},\\
   \nonumber
   \Gamma_1=\beta+\frac{(4-3\beta)^2(\gamma-1)}{\beta+12(\gamma-1)(1-\beta)},\\
   \beta=\frac{\rho kT/\mu m_p}{\bar{a} T^4/3+\rho k T/\mu m_p}.
   \label{gambet}
   \end{equation}
   Here, $\beta$ (ratio of gas pressure to total pressure) close to $0$ for radiation 
   dominated flow (highly relativistic flow) and close to $1$ for
   gas dominated flow.
Now combining (\ref{ec}),(\ref{rmom}), (\ref{azmom}) and (\ref{enteq}), we get
\begin{equation}
\frac{dv}{dx}=\frac{f_1(x,v,c_s)}{f_2(v,c_s)},
\label{dvdx}
\end{equation}
where,
\begin{eqnarray}
\nonumber
f_1(x,v,c_s)&=&\left[c_s^2\left(\frac{3}{2x}-\frac{1}{2F}\frac{dF}{dx}\right)-\gamma\left(F-\frac{\lambda^2}{x^3}\right)\right]
\left[v(\Gamma_1+1)-\frac{4\alpha\Lambda}{v(3\gamma-1)}\right] \\
&-&\left[\frac{2\alpha c_s^2}{(3\gamma-1)xv}+
\frac{\alpha v}{x}-\frac{2\lambda}{x^2}\right]\Lambda+\left(\frac{3}{2x}-\frac{1}{2F}\frac{dF}{dx}\right)
v c_s^2(\Gamma_1-1),
\label{f1}
\end{eqnarray}
\begin{eqnarray}
f_2(v,c_s)=\left[1-\frac{2c_s^2}{(3\gamma-1)v^2}\right]\alpha\Lambda-c_s^2(\Gamma_1-1)
+\left(\gamma v-\frac{c_s^2}{v}\right)\left[v(\Gamma_1+1)-\frac{4\alpha\Lambda}{(3\gamma-1)v}\right]
\label{f2}
\end{eqnarray}
and $\Lambda=\gamma(\Gamma_3-1)f\alpha\left(\frac{I_{n+1}}{I_n}\frac{c_s^2}{\gamma}+v^2\right)$.\\
As far away from the black hole, $v<c_s$ and close to it, $v>c_s$, there must be an intermediate
location, where the denominator of (\ref{dvdx}) must be vanished. To have a smooth solution, at that
location, numerator has to be zero. This location is called the sonic point or critical point ($x_c$). The existence of 
sonic location plays an important role in accretion phenomena. Although, in case of accretion flow of matter
around black hole sonic point must exist, for that around GOHS (say, neutron star) it is not always necessary
(Mukhopadhyay 2002a). When shock forms in an accretion disk around a neutron star (Chakrabarti \& Sahu 1997, 
Mukhopadhyay 2002a), sonic points play an important role. From the global study of the sonic points in an accretion disk, one can 
understand about the stability of physical parameter region, which we will discuss in the next section. 

Now, when the sonic point exists in an accretion disk around GOHS, $f_2(v_c,c_{sc})=0$ at $x=x_c$. Thus
at that location Mach number can be written as
\begin{eqnarray}
M_c^2=\frac{{\cal B}+\sqrt{{\cal B}^2-4\cal{A}\cal{C}}}{2\cal{A}},
\label{mac}
\end{eqnarray}
where
\begin{eqnarray}
\nonumber
{\cal A}&=&\alpha^2\gamma f(\Gamma_3-1)+\gamma(\Gamma_1+1)-\frac{4\alpha^2f(\Gamma_3-1)\gamma^2}{3\gamma-1},\\
\nonumber
{\cal B}&=&2\Gamma_1-\frac{4\alpha^2\gamma f(\Gamma_3-1)}{3\gamma-1}\left(1-\frac{I_{n+1}}{I_n}\right),\\
{\cal C}&=&\frac{2\alpha^2 f(\Gamma_3-1)}{3\gamma-1}\frac{I_{n+1}}{I_n}.
\end{eqnarray}
Subsequently, at $x=x_c$, $f_1(x_c,v_c,c_{sc})=0$. Thus, using (\ref{mac}) we can eliminate $v_c$ from $f_1(x_c,v_c,c_{sc})=0$
and get an algebraic equation for $c_{sc}$ and $x_c$, which we can solve to find out sound speed at sonic location.
To find out $\frac{dv}{dx}|_c$, we apply l'Hospital's rule to (\ref{dvdx}).
In order to understand the fluid properties in accretion disk, we have to solve (\ref{dvdx}) 
with an appropriate boundary condition. Also, integrating (\ref{azmom}), we get the angular momentum of the 
accreting matter as
\begin{eqnarray}
\lambda=\frac{\alpha x c_s^2}{v}\left(\frac{2}{3\gamma-1}+M^2\right)+\lambda_{in},
\end{eqnarray}
where $M$ is the Mach number of the flow and $\lambda_{in}$ is the angular momentum at the
inner edge of accretion disk.

Now integrating (\ref{ec}) and (\ref{rmom}), we can write the entropy and energy of the flow at sonic point as 
\begin{equation}
\dot{\cal M}_c=x_c^{3/2} F_c^{-1/2}(\gamma+1)^{q/2}\left(\frac{\frac{\lambda^2}{x_c^3}-F_c}{\frac{1}{F_c}
\frac{dF}{dx}|_c-\frac{3}{x_c}}\right)^{\frac{\gamma}{\gamma-1}}
\label{muc}
\end{equation}
and
\begin{equation}
E_c=\frac{2\gamma}{(\gamma-1)}\left(\frac{\frac{\lambda^2}{x_c^3}-F_c}{\frac{1}{F_c}\frac{dF_c}{dx}|_c
-\frac{3}{x_c}}\right)+V_c+\frac{\lambda^2}{2x_c^2},
\label{Ec}
\end{equation}
where $V_c=(\int F dx)|_c$ and $q=\frac{(\gamma+1)}{2(\gamma-1)}$. It can be mentioned that,
disk entropy and accretion rate are related by a simple relation as $\dot{\cal M}=(\gamma K)^n\dot{M}$.
While $\dot{M}$ is a conserved quantity for the particular accretion flow, $\dot{\cal M}$ is not,
as it contains $K^n$ which carries the entropy information that is not conserved in a dissipative system. 
As a boundary condition for the particular flow, we have to supply the sonic energy $E_c$. 
Then from (\ref{Ec}), we can find out the sonic location $x_c$. Therefore, knowing $x_c$ one can
easily find out the fluid velocity and sound speed at the sonic point from $f_1=f_2=0$ with the 
help of (\ref{mac}) for a particular accretion flow. These have to be supplied as further boundary 
conditions of the flow.

Now let us come to the issue of the formation of shock. Mukhopadhyay (2002a) showed that in accretion 
disk around neutron star, double shock is very natural in certain physical situations.
Here, whenever we mention the shock, we will mean Rankine-Hugoniot shock (Landau \& Lifshitz 1987).
If we generalise the conditions to form a shock in an accretion disk given by Chakrabarti (1989) for
rotating GOHS, we get
\begin{equation}
\frac{1}{2}M_+^2 c_{s+}^2+nc_{s+}^2=\frac{1}{2}M_-^2 c_{s-}^2+nc_{s-}^2,
\label{scke}
\end{equation}
\begin{equation}
\frac{c_{s+}^\nu}{\dot{\cal M}_+}\left(\frac{2\gamma}{3\gamma-1}+\gamma M_+^2\right)
=\frac{c_{s-}^\nu}{\dot{\cal M}_-}\left(\frac{2\gamma}{3\gamma-1}+\gamma M_-^2\right),
\label{sckmom}
\end{equation}
\begin{equation}
\dot{\cal M}_+>\dot{\cal M}_-,
\label{sckent}
\end{equation}
where
\begin{equation}
\dot{\cal M}=Mc_s^{2(n+1)}\frac{x_s^{3/2}}{\sqrt{F_s}}.
\label{sckent1}
\end{equation}
Here, subscript '$-$' and '$+$' indicate the quantities just before and after the shock 
respectively, $x_s$ indicates the shock location and $\nu=\frac{3\gamma-1}{\gamma-1}$.
From (\ref{scke})-(\ref{sckent1}), it is very clear that except the entropy expression, all remain unchanged
with respect to the case of a non-rotating central object. Also from (\ref{sckmom}) and (\ref{sckent1}), we get
the shock invariant quantity as
\begin{equation}
C=\frac{\left(\frac{2}{M_+}+(3\gamma-1)M_+\right)^2}{M_+^2(\gamma-1)+2}=
\frac{\left(\frac{2}{M_-}+(3\gamma-1)M_-\right)^2}{M_-^2(\gamma-1)+2},
\label{sckinv}
\end{equation}
which remains unchanged with respect to a non-rotating case. If only all the conditions, 
(\ref{scke})-(\ref{sckent}) and (\ref{sckinv}) are simultaneously satisfied by the matter, shock will
be generated in an accretion disk.

\section{Global analysis of the parameter space}

One of our aim is to check, how does the rotation of GOHS affect the disk parameter region known
for non-rotating case. More precisely, we will check, how does the rotation of GOHS affect the sonic 
location, structure as well as the stability of disk. As mentioned in \S 1, we will follow the methodology
adopted by Chakrabarti (1996b) for global solution and Mukhopadhyay \& Chakrabarti (2001) for stability
analysis. In Fig. 1, we show the variation of disk entropy as a function of sonic location.
The intersection to the horizontal line (which indicates a constant entropy curve) with all the
curves indicate the sonic points of the accretion disk for that particular entropy and rotation
to the GOHS. It is clearly seen that at a particular entropy, if the co-rotation of GOHS
increases, the inner sonic points shift to more inner edge of the disk, which are comparatively unstable region
(because, beyond of $x_s$, disk orbit is not stable and decrement of the value of $x_s$ is slower than
that of $x_c$ with $J$). For $J=0.5$, the inner sonic point disappears, as we see that dashed curve
terminates on the radial axis at $x_c\sim 6$ and it does not turn up to give rise the inner sonic point.
The reason behind it is that, for higher the co-rotation, the angular momentum of the system will
become higher and radial matter speed will be unable to overcome the centrifugal barrier close to the
GOHS and that will make the disk unstable at the inner edge. For the counter-rotating cases, if 
$J$ increases in magnitude for a particular entropy, disk will become more stable, as the radial speed will be able to
overcome the centrifugal barrier at a larger radius compared to a co-rotating case and matter may attain supersonic speed at a larger distance, which
is more stable region of the disk. However, for a high counter rotation, angular momentum of the system
will become too low and the accretion will tend to 'Bondi-like flow' (Bondi 1952) with the existence of single
sonic point. 

It is seen that for $\alpha=0-0.1$, the variation of curves are almost same and only
significant changes come into the picture for $\alpha >0.1$. Also for the lower viscosity, if the co-rotation 
is high ($J$ is high), the possibility to have the all three sonic points in the disk is high. 
As the viscosity increases, the possibility for the existence of all three sonic points diminishes.
In order to preserve the existence of sonic point for the high viscous cases, $J$ must reduce to zero and
subsequently, more negative.
As a whole, from Fig. 1a-c we can conclude that for lower $\alpha$, multiple sonic points may exist, but for high 
$\alpha$, there is only one sonic point which is inner one in the accretion disk.
Similar features come out from Fig. 2 but in terms of energy of the sonic point in accretion disk.
As we know that the observed angular frequency for the candidate 4U 1636-53 is 582Hz, which is equivalent
to $J=0.2877$, one of the value of $J$ chosen for our discussion is $0.2877$.
For both Fig. 1 and 2, sonic points with negative slope of the curve indicate the
locations of 'saddle-type' or 'X-type' sonic point and positive slopes indicate the 'centre-type' 
or 'O-type' sonic point. Thus, the rotation of GOHS and viscosity of the disk are the important factors to the formation and 
location of sonic points, which are related to the structure of accretion disk.

\begin{figure}
\psfig{file=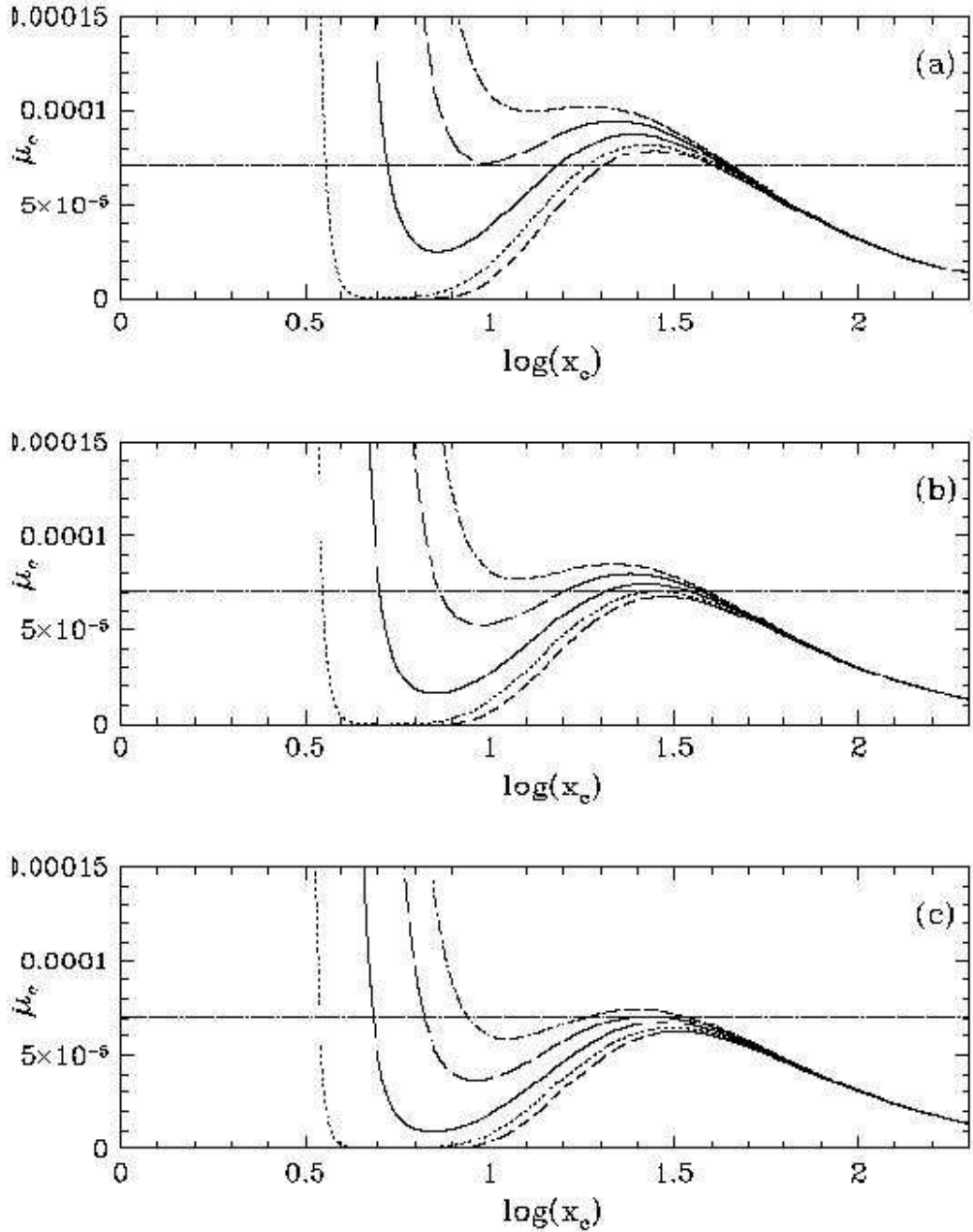,width=0.9\textwidth} 
\caption{ Variation of sonic entropy as a function of sonic location for various values of 
$J$ when (a) $\alpha=0,f=0$, (b) $\alpha=0.4,f=0.5$, (c) $\alpha=0.8,f=0.5$. 
Central solid curve indicates non-rotating case ($J=0$), while the dotted and dashed curves are 
for $J=0.2877, 0.5$ respectively and long-dashed and dot-dashed curves are for $J=-0.2877, -0.5$ respectively. 
The horizontal line indicates the curve of constant entropy of $7\times 10^{-5}$. 
The other parameters are $\lambda=3.3$, $\gamma=4/3$. 
} \label{fig1} 
\end{figure}

\begin{figure}
\psfig{file=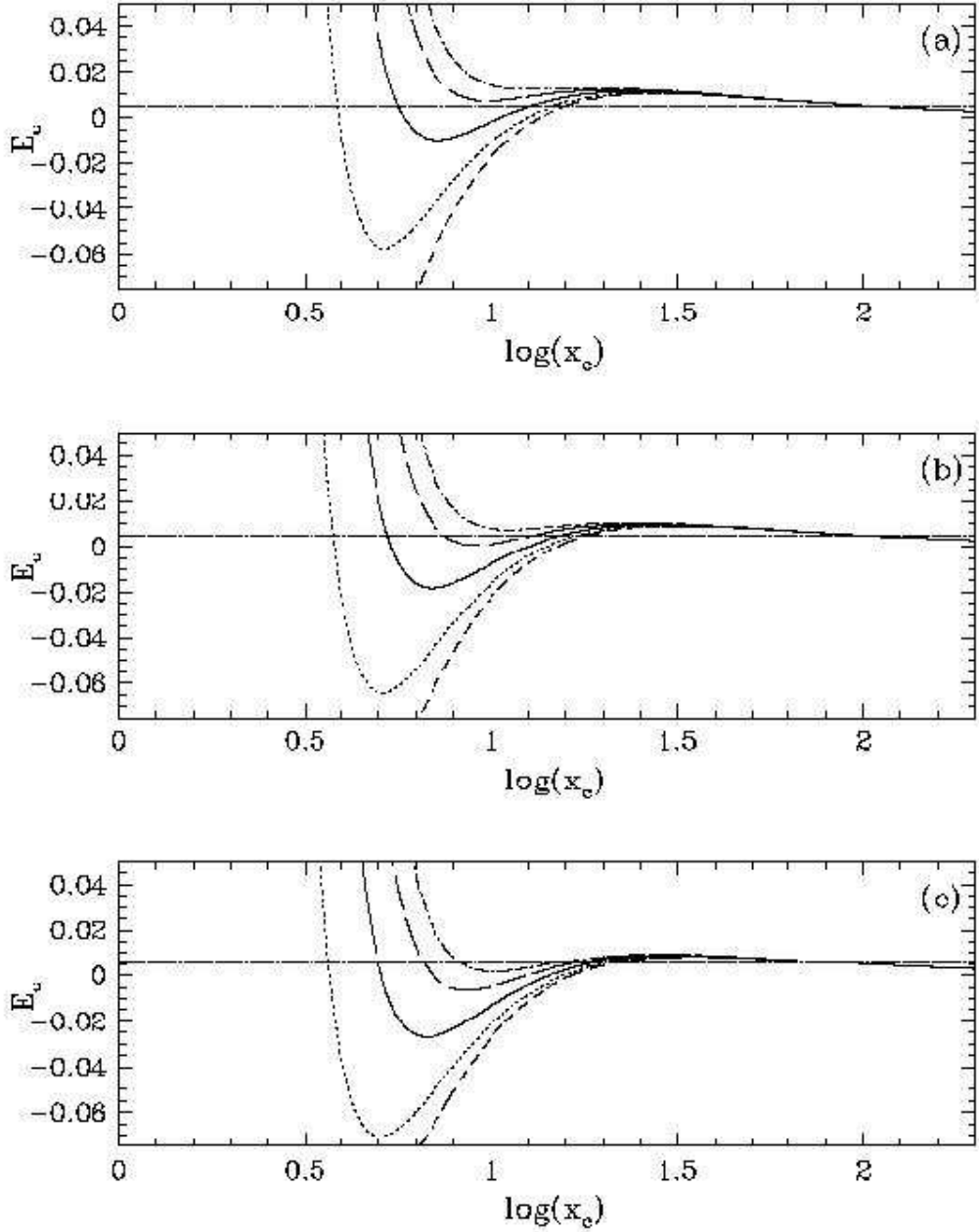,width=0.9\textwidth} 
\caption{
Same as \ref{fig1}, but sonic energy is plotted in place of entropy. The horizontal line indicates 
the curve of constant energy of $0.005$.}
\label{fig2}
\end{figure}

\begin{figure}
\psfig{file=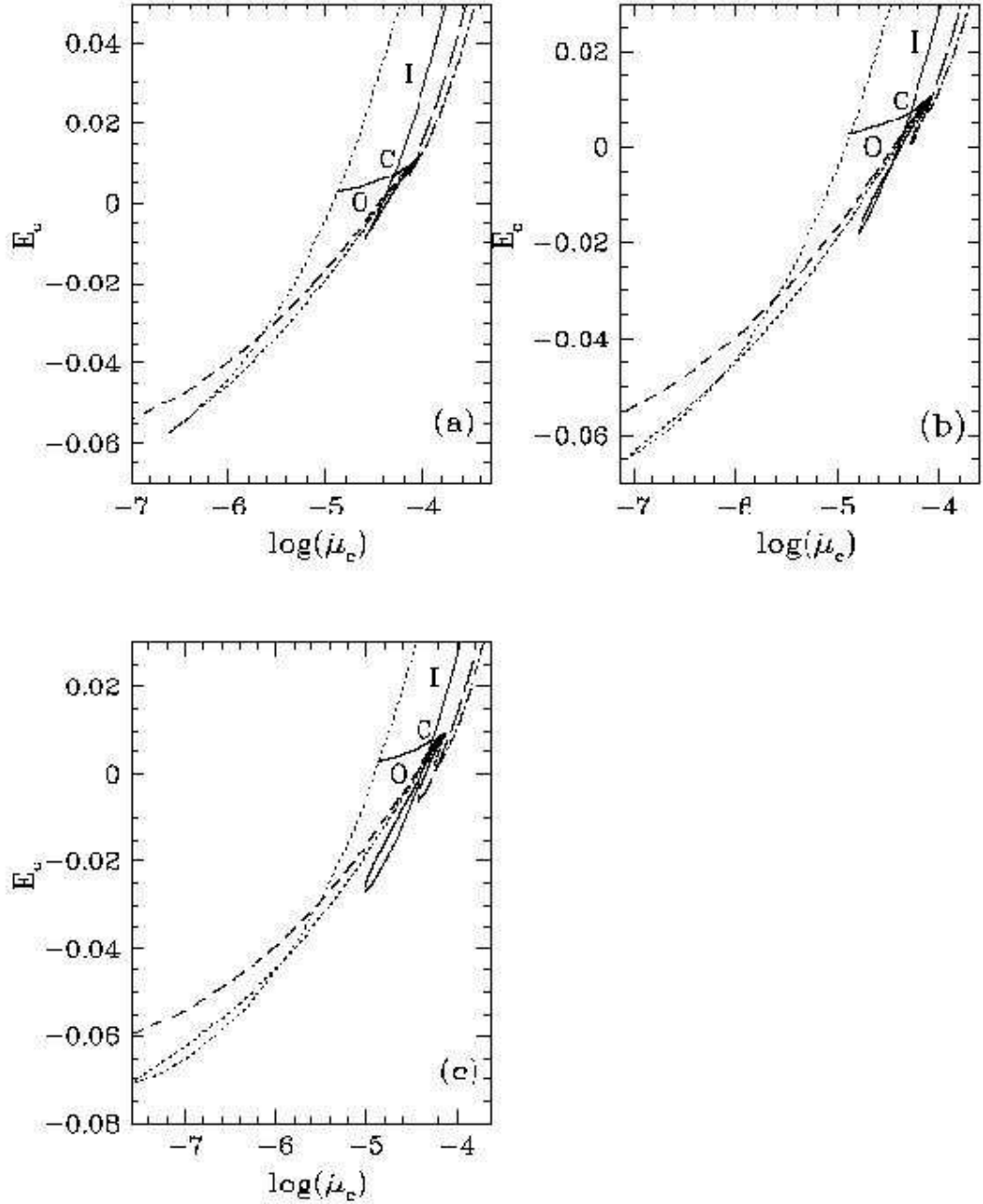,width=0.9\textwidth} 
\caption{ Variation of sonic energy as a function of sonic entropy for various values of 
$J$ when (a) $\alpha=0,f=0$, (b) $\alpha=0.4,f=0.5$, (c) $\alpha=0.8,f=0.5$. 
Solid curve indicates non-rotating case ($J=0$), while the dotted and dashed curves are 
for $J=0.2877, 0.5$ respectively and long-dashed and dot-dashed curves are for $J=-0.2877, -0.5$ respectively. 
O and I indicate the locus of outer and inner sonic point respectively, while C is the intersection
of outer and inner sonic point locus. The other parameters are $\lambda=3.3$, $\gamma=4/3$. 
}\label{fig3}
\end{figure}

\begin{figure}
\psfig{file=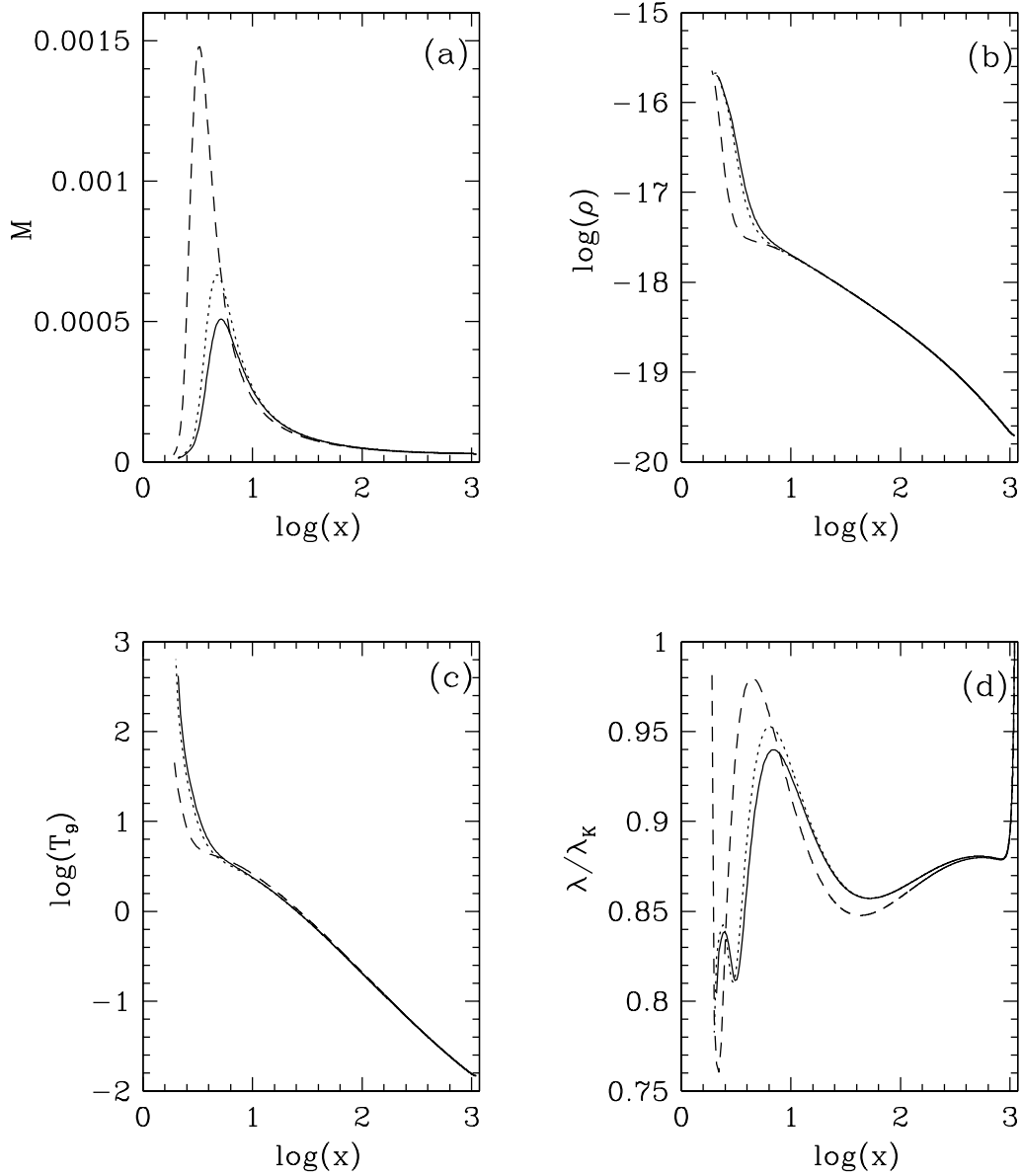,width=0.9\textwidth} 
\caption{
Variation of (a) Mach number, (b) density, (c) temperature in unit of $10^9$, (d) ratio of sub-Keplerian
to Keplerian angular momentum of the accreting fluid as a function of radial coordinate. Solid, 
dotted and dashed curves are for $J=0,0.1,0.5$ respectively. Other parameters are $\alpha=0.0001, f=0.1,
M=2, \dot{M}=2, \beta=0.03$.
}\label{fig4}
\end{figure}

\begin{figure}
\psfig{file=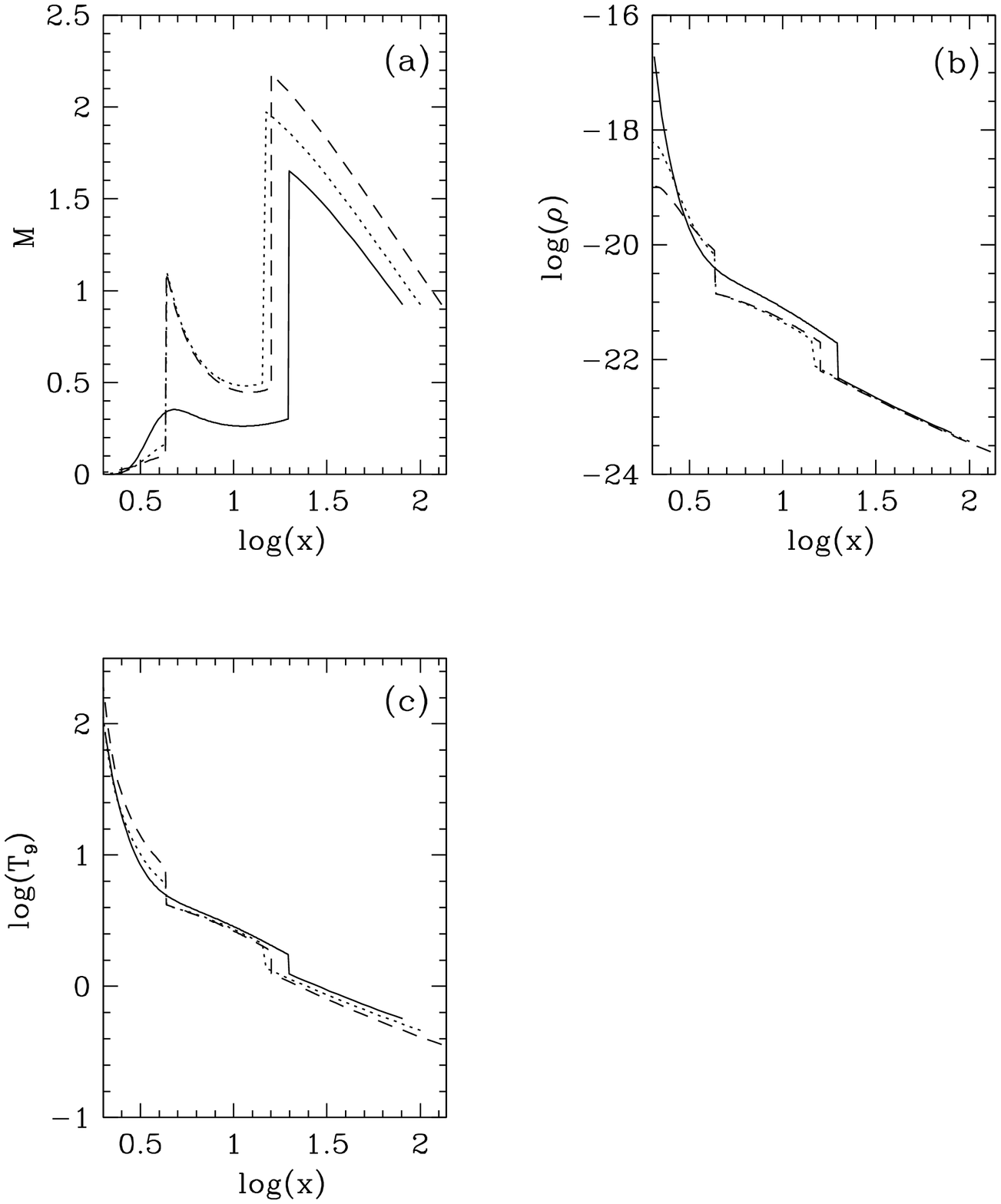,width=0.9\textwidth} 
\caption{
Variation of (a) Mach number, (b) density, (c) temperature in unit of $10^9$
as a function of radial coordinate for the accretion disk around 4U 1636-53. Solid, dotted and dashed curves are respectively for 
(i) $\alpha=0, f=0$, (ii) $\alpha=0.02, f=0.2$, (iii) $\alpha=0.05, f=0.5$. 
Other parameters for 4U 1636-53 are $J=0.2877, \lambda_c=3, M=1.4, \dot{M}=1, \beta=0.03$.
}
\label{fig5}
\end{figure}

Figure 3 indicates the variation of energy with entropy at sonic locations, when $J$ is parameter 
for different viscosity of the fluid. It is seen that, with the increase of $J$ (co-rotation), inner sonic
points of the disk shift towards the lower entropy region keeping the outer sonic points unchanged
and the disk becomes unstable. We know that the shock in accretion disk can be formed only if there is a 
possibility of matter in the outer sonic point branch
of lower entropy to jump to the inner sonic point branch of higher entropy. As because, with 
the increase of $J$, inner sonic point branch itself shifts to the lower entropy
region which is comparatively unstable, the shock as well as disk also become unstable. 
For $J=0.5$, the inner sonic point branch disappears totally and only outer and O-type
sonic point branches remain. 
Clearly, for a highly rotating GOHS, the possibility of formation of shock reduces as 
well as the shock becomes unstable, as the inner sonic point branch itself disappears
or tends to disappear. The physical reason behind it is that, with the increase of value of $J$,
the angular momentum of the system increases, which helps the disk to maintain
a high azimuthal speed of matter upto a very close radius to the GOHS, as a result the radial speed 
of the matter can overcome the corresponding sound
speed at a very inner radius only. On the otherhand, the inner edge of the accretion disk is comparatively
less stable, as the entropy decreases at lower radii. As the sonic points form at a more inner radii, 
with the increase of $J$, disk tends to an unstable situation. 

In case of the counter-rotation of GOHS, as $J$ increases in magnitude, 
the inner sonic point branch shifts towards a higher entropy region, and the disk as well as the
shock become more stable. The physical reason to have a more stable inner sonic
point branch for the counter-rotating case is the following. 
As the counter rotation of central object increases, that is the angular momentum of the system
reduces, the matter falls freely towards GOHS at larger radii and it becomes supersonic
at a comparatively outer edge of the disk. Thus, the disk as well as shock become more stable.
However, for a very high counter-rotation of GOHS, this inner sonic point branch itself tends to merge 
to the outer sonic point branch, and the entire system shift towards the outer side which is more stable. As in case of 
higher counter-rotation, the angular momentum of the system becomes very small, the corresponding
centrifugal pressure onto the matter becomes insignificant, as a result the possibility of shock 
diminishes again. Thus, if the angular momentum of GOHS
increases or decreases significantly, the shock wave in accretion disk becomes unstable and the disk itself
tends to an unstable situation. In Fig. 3, 'C' is the point, where the flow can pass through the inner and outer sonic point 
simultaneously. The parameters around C are important to study the shock in accretion disk. 

As the viscosity of flow increases, the possibility of intersection between inner and outer sonic
point branch decreases. With the increase of viscosity, more regions
containing the X-type sonic points become O-type. The outer X-type sonic points recedes in further out
and inner one proceeds in further inwards. Also for higher viscosity, the outer sonic points may no longer
remain as X-type and only the inner sonic points may exist. Thus, the possibility of shock will reduce.

\section{Fluid Properties of Accretion Disk }

Now we will discuss the behaviour of viscous accreting fluid around rotating GOHS. 
In the previous section, we have already confirmed that the rotation significantly affects the 
global parameter space of accretion disk. Here, our intention is to see, how the various fluid dynamical
results of an accretion disk are affected for various values of $J$ and $\alpha$. For a certain choice of
physical parameter set, if the existence of sonic radii are possible in the accretion disk around GOHS,
the flow structures can be studied for the corresponding choice of sonic radii (or energy) of the accretion flow.

In Fig. 4a, we show the variation of Mach number for three different values of $J$ at a particular viscosity. As the
viscosity is low, the rate of energy momentum transfer in matter is less. It is known that for an accretion around
neutron star, in a certain cases, matter may be always subsonic 
(Chakrabarti \& Sahu 1997; Mukhopadhyay 2002a) and at the stellar surface
matter speed reduces to zero. Owing to this fact, here such a situation is considered where 
the speed of the accreting fluid is low, which accounts for the 
high residence time of matter in the disk. The angular momentum of the disk is high at $x\sim 10$ (see Fig. 4d)
and thus radial matter speed is low. Subsequently, though the disk angular momentum goes down, as the matter comes closer to the
stellar surface, it starts to decelerate and the radial matter speed still remains low.
This results to a higher the possibility of cooling, and thus cooling factor
$f$ is chosen as $0.1$. As the energy momentum transfer rate of matter is low, the accretion rate is chosen 
intermediate. We have chosen a standard mass of GOHS as $2M_\odot$. If the GOHS
is chosen to be non-rotating, angular momentum of the system is less and the centrifugal barrier is minimum. 
With the increase of $J$, this centrifugal barrier increases, and it becomes very high particularly 
for $J=0.5$. Mach number profiles clearly indicate that at the surface of GOHS, accreting fluid stops. Also from Fig. 4b and 4c,
it comes out that sudden deceleration of the accreting fluid gives rise to a sudden enhancement of density
and temperature close to GOHS. Here, the temperature is thought to be cooled down by a factor of $1/30$
through the inverse-Compton effect in the radiation pressure dominated relativistic flow. Thus, we choose $\beta\sim 0.03$.
Nevertheless, the temperature is still very high, which was also reported earlier (Mukhopadhyay 2002a) in the
context of non-rotating neutron star. The density profiles are plotted in unit of $\frac{c^6}{G^3M^2}$. 
Figure 4d shows the ratio of sub-Keplerian to Keplerian angular momentum variation of the accreting
fluid as a function of radial coordinate. It is seen that, at the boundary between Keplerian and 
sub-Keplerian disk, $\lambda_k/\lambda\rightarrow 1$, and from that radius ($x_K$) we start our calculation
as our discussion is for sub-Keplerian accretion disk. From the $\lambda_k/\lambda$ profile, it is
again clear that for higher $J$, centrifugal barrier becomes stronger. Far away from GOHS, the effect of centrifugal
barrier is weak, but as the matter approaches towards GOHS, it starts to dominate and at $x\sim 10$ it
becomes stronger. Then again it goes off. However, at very close to the GOHS, again this centrifugal
effect comes into the picture as the matter falls into the strong rotational field of GOHS. This 
effect appears very prominently for high value of $J$, which is here chosen to be $0.5$. Actually, this effect comes into the
picture, once the matter enters into the corresponding ergosphere of HT geometry. But the important point to be noted that if the
outer radius of GOHS is greater than the radius of ergosphere, this feature is not possible to occur. 

In Fig. 5, we show the results of accretion phenomena for 4U 1636-53, whose angular momentum $J=0.2877$. 
The examples are shown, where the sonic point(s) exist in the flow and shock forms in the accretion disk 
around 4U 1636-53 for different 
viscosity parameters. Figure 5a shows the variation of Mach numbers as a function of radial coordinate.
It reflects that, for a small $\alpha$ ($\sim 0$) only shock is possible at $x=19.02$ in the accretion disk.
As $\alpha$ increases, the energy momentum transfer rate increases, rate of infalling the matter enhances and 
velocity of the accreting fluid becomes high at the inner edge of the disk. But, because of the inner
hard surface, matter has to stop at close to GOHS and naturally another shock forms at the inner edge
of the accretion disk. The shock locations for $\alpha=0.02$ and $0.05$, are at $x=14.27, 4.38$ and 
$x=15.37, 4.31$ respectively. If the viscosity is high, the cooling factor is intermediate (see Fig. 5 caption). 
Also, the angular momentum of matter at the sonic point is chosen as, $\lambda_c=3$, and for the radiation dominated, 
inverse-Comptonised flow, $\beta$ is chosen to be $0.03$. Figures 5b and 5c show the variations of density and temperature.
The regions, where velocity decelerates abruptly, temperature and density also jumps up in a significant manner.
Here also, the inverse-Comptonised temperature is very high as the cases of Fig. 4.


\section{Summary and Discussion}

Here, we have studied the viscous accretion phenomena around rotating gravitational object with 
hard surface (GOHS), i.e., mostly around compact object like neutron star and strange star. We started
with a proposal of a new pseudo-Newtonian potential by means of Hartle-Thorne geometry, as this 
metric can describe the space-time of interior as well as exterior to a slowly rotating
star. We have shown that this potential can describe all the essential relativistic properties 
like, radius of marginally bound and stable orbit, specific mechanical energy etc., of accretion disk
within $10\%$ error. Then, along with this potential, we have described the accretion disk by
using non-relativistic equations. Therefore, we can say, our method is a 'pseudo-general-relativistic'. 
Once the pseudo-Newtonian potential is proposed, we have applied it to the study of relativistic
fluid properties of accretion disk around rotating GOHS. We have prescribed the set of basic equations
for vertically averaged, thin, viscous accretion disk. Then, we have analysed the viscous parameter
space globally. Subsequently, we have discussed about the properties of viscous fluid and the effect
of rotation on the fluid behaviour. Examples are shown for the accretion disk around an observed candidate 4U 1636-53,
which is one of the fast rotating compact object with angular frequency 582Hz (equivalent $J=0.2877$).
As starting from a new pseudo-potential, we have discussed upto the solution of viscous accretion flow, this
is one of the most self consistent paper on accretion disk.

We have studied the stability of accretion disk. Most of the earlier analysis of structure and
stability of the accretion disk have been done around non-rotating central object. As no cosmic
object is static, it is very important to consider the effect of rotation, particularly
for the discussion of an inner edge of the disk. We have found that, when the rotation to a GOHS
is incorporated, the locations of sonic point (if any) get shifted to the unstable region 
and the shock (if any) gets affected in the disk, and thus the disk structure gets influenced. 
Therefore, we can conclude that the rotation should be incorporated
in the study of an accretion disk. Any related conclusion is dependent on the rotation parameter of 
GOHS. The mentioned disk properties and phenomena not only get modified and/or shifted in location, 
sometimes disappeared completely. Prasanna \& Mukhopadhyay (2003), worked on the stability 
analysis of accretion disk around rotating compact objects in a perturbative manner. They incorporated the 
rotational effect of central object in a disk indirectly by the inclusion of Coriolis acceleration term. 
Here, this rotational effect is brought directly from the metric itself.

From the global analysis of sonic points, we have seen that for a higher co-rotation of GOHS,
the disk becomes unstable for a particular angular momentum of the accreting matter. As the formation
of shock needs a stable inner sonic point in the accretion disk, for the higher value of $J$, shock is unstable
as the inner region of disk is unstable. On the other hand,
for counter-rotating cases, the angular momentum of system reduces and the matter falls more 
strongly to a GOHS. As there is no significant centrifugal barrier to slow
down the matter in disk, the possibility of shock reduces again. Also the branch of inner sonic point
merges or tends to merge to that of outer sonic point. Thus, we can conclude that 
the parameter region, where the shock is expected to form for non-rotating
GOHS (Chakrabarti \& Sahu 1997), is affected for rotating ones and the possibility of shock is reduced for both the co-rotating 
and counter-rotating GOHS. Also for the higher viscosity, both the inner and outer
sonic point branch merge each other and the possibility of shock reduces again. 
Actually, with the increase of viscosity, the region containing the X-type sonic points tends to become O-type. 
The outer X-type sonic points recede in more outwards and inner ones proceed in inwards more. 

As the incoming matter slows down abruptly at the surface of GOHS as well as at two
shock locations (if two shocks form), the overall density becomes higher
for an accretion disk around GOHS compared to that for a black hole. Similarly, the temperature in the
disk is higher for a GOHS. Thus, one can conclude that the accretion disk around a GOHS is very favourable
for nucleosynthesis. We know that the accretion disk around a black hole is enough hot for nucleosynthesis
(Mukhopadhyay 1999; Mukhopadhyay \& Chakrabarti 2000), which are different from that in star.
As the density and temperature may be much higher around a GOHS, 
more efficient nucleosynthesis is expected particularly at an inner region of the disk. In 
early, we saw that the high 
temperature of accretion disk around a black hole is very favourable for the photo-dissociations and the proton
capture reactions (Chakrabarti \& Mukhopadhyay 1999). As the accretion disk around a GOHS is hotter
than that around a black hole, even $^4\!He$ which has a high binding energy,
may dissociate into deuterium and then
into proton and neutron. If we consider the accreting matter to come
from the nearby 'Sun like' companion star, the initial abundance of
$^4\!He$ in the accreting matter is supposed to be about $25\%$. Therefore, by the dissociation of
this $^4\!He$, neutron may produce in a large scale, which could give rise to the
neutron rich elements. Guessoum \& Kazanas (1999) showed that the profuse
neutron may be produced in the accretion disk and through the spallation
reactions {\it lithium} may be produced in the atmosphere of the star.
When the neutron comes out from the accretion disk by the formation of
an outflow, in that comparatively cold environment,
$^7\!Li$ may be produced which can be detected on the stellar surface.
In early, it was shown that the metalicity of the galaxy may be influenced
when outflows form in the hot accretion disk around black holes (Mukhopadhyay \& Chakrabarti 2000).
In case of the lighter galaxy, the average abundance of the isotopes of
$Ca$, $Cr$ may significantly change.
Also the abundance of lighter elements, like the isotopes of $C$, $O$,
$Ne$, $Si$ etc. may increase significantly. As the
temperature of the accretion disk around GOHS is higher, the expected
change of abundance of these elements and the corresponding influence on
the metalicity of the galaxy is expected to be high.

We know that, out of an observed pair of kilohertz QPO frequencies for a particular
candidate, one is the oscillation due to its Keplerian motion (Osherovich \& Titarchuk 1999). For 4U 1636-53,
lower frequency is 950Hz (Wijnands et al. 1997) which may be the Keplerian one. 
According to our pseudo-potential, if we calculate this Keplerian frequency for 4U 1636-53, 
it comes out to be 1027Hz for $M=1.4M_\odot$ and $x_K=18$, which is not much over estimated
with respect to the observed one. Thus, we can give a theoretical prediction of QPO, at least for
one, out of a pair. Similarly, for other candidates, one can calculate Keplerian frequency
according to our potential and compare with observation. The detailed theoretical study of QPO
based on this present scheme will be pursued elsewhere.

\section*{Acknowledgments} 

SG is thankful to Inter-University Centre for Astronomy and Astrophysics for Summer Training Programme, 2002,
where this work was initiated.


\end{document}